\documentclass[twocolumn,showpacs,preprintnumbers,amsmath,amssymb]{revtex4}
\usepackage{epsfig}
\usepackage{graphicx}% Include figure files
\usepackage{dcolumn}% Align table columns on decimal point
\usepackage{bm}% bold math

\newcommand{\om}{\omega}
\newcommand{\Om}{\Omega}

\newcommand{\be}{\begin{equation}}
\newcommand{\ee}{\end{equation}}

\begin{document}

\preprint{APS/123-QED}

\title{Atom Interferometry tests of the isotropy of post-Newtonian gravity}% Force line breaks with \\

\author{Holger M\"uller$^1$}
\email{holgerm@stanford.edu}
\author{Sheng-wey Chiow$^1$, Sven Herrmann$^1$}
\author{Steven Chu$^{1,2}$}
\affiliation{$^1$Physics Department, 382 Via Pueblo Mall,
Stanford, California 94305, U.S.A. \\ $^2$Lawrence Berkeley
National Laboratory and Department of Physics, University of
California, Berkeley, 1 Cyclotron Road, Berkeley, California
94720, U.S.A.}
%\author{Achim Peters}
%\affiliation{Institut f\"ur Physik, Humboldt-Universit\"at zu Berlin, Hausvogteiplatz 5-7, 10117 Berlin, Germany}
\author{Keng-Yeow Chung}
\affiliation{Physics Dept., National University of Singapore, 2
Science Drive 3, Singapore 117542}

\date{\today}% It is always \today, today,

\begin{abstract}
We present a test of the local Lorentz invariance of
post-Newtonian gravity by monitoring Earth's gravity with a
Mach-Zehnder atom interferometer that features a resolution of
about $8\times 10^{-9}$\,g/$\sqrt{\rm Hz}$, the highest reported
thus far. Expressed within the standard model extension (SME) or
Nordtvedt's anisotropic universe model, the analysis limits four
coefficients describing anisotropic gravity at the ppb level and
three others, for the first time, at the 10ppm level. Using the
SME we explicitly demonstrate how the experiment actually compares
the isotropy of gravity and electromagnetism.
\end{abstract}

\pacs{03.75.Dg, 11.30.Cp, 11.30.Qc, 04.25.Nx}% PACS, the Physics and Astronomy
\maketitle

%\section{Introduction}

The description of gravitation by a dynamic geometry of
space-time, Einstein's general relativity (GR), is based on the
Einstein equivalence principle. This encompasses the universality
of free fall (UFF), local position invariance (LPI), and local
Lorentz invariance (LLI), which also underlies the
non-gravitational standard model of particle physics. Attempts to
unify GR and the standard model have failed so far. This suggests
that one of their foundations might be violated at some level of
precision \cite{AmelinoCamelia}. So far, tests of the UFF and LPI
have not identified violations \cite{AmelinoCamelia}. LLI has been
tested experimentally for many sectors of the standard model, such
as for photons ('Maxwell sector'), electrons, protons, and
neutrons \cite{AmelinoCamelia,Kostelecky,Mattingly}. No Lorentz
violation has been identified, although the coverage of parameter
space is still incomplete. Far less attention, however, has been
paid to the LLI of the gravitational (`Einstein') sector, in spite
of the pioneering work of Nordtvedt and Will in the 1970ies.
Motivated by that fact that anisotropies arise in various theories
of gravity other than GR \cite{Nordtvedt76}, they have ruled out a
Lorentz-violating anisotropy in gravity by searching for an
anomalous time-dependence of the acceleration of free fall $g$ on
Earth \cite{Will71,NordtvedtWill72,Nordtvedt76}.

The success of GR and the standard model implies that any Lorentz
violations are tiny. This and the relative weakness of gravity
means that only exceptionally sensitive experiments can hope to
detect Lorentz violation in gravity. A relatively recent addition
to these is precision atom interferometry
\cite{KasevichChu,Chureview}. This has been serving, for example,
in measurements of the fine structure constant \cite{hm}, $g$
\cite{Peters} and its gradient \cite{Snaden98}, the Sagnac effect
\cite{Gustavson}, and Newton's constant $G$ \cite{Fixler} with
sensitivities that compare favorably with other state-of-the-art
instruments. One reason for its outstanding precision is that the
motion of neutral atoms can realize a freely falling frame to a
high accuracy and that this motion can be interrogated by laser
radiation in a tremendously precise way. As a result, tests of
post-Newtonian gravity with atom interferometry have been proposed
that could rival or exceed the precision of classical ones
\cite{Dimopoulos}.

Here, we report on a first step in this direction: We describe the
highest resolution atomic gravimeter reported thus far
\cite{Kengyeow}. We then analyze the influence of Lorentz
violation in gravity. By explicitly including possible Lorentz
violation in electrodynamics, we explicitly show how this (like
any) isotropy test is actually a comparison of two sectors.
Finally, we report a test of the LLI of
post-Newtonian gravity by testing its isotropy.\\

%\section{Experiment}
%\subsection{Principle}

\begin{figure}
\centering \epsfig{file=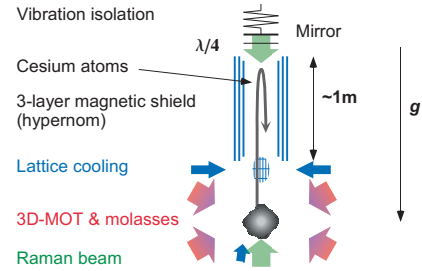, width=0.3\textwidth}
\caption{\label{setup} Setup. $\lambda/4$; $1/4$-wave retardation
plate. Vibrations of the top mirror are reduced to below $5\times
10^{-9}\,g/\sqrt{\rm Hz}$ in a frequency range of 0.1-10\,Hz by a
sophisticated active vibration isolator \cite{Hensleyvib}.}
\end{figure}

Our experimental setup (Fig. \ref{setup}) assembles about $10^9$
Cs atoms within 650\,ms from a background vapor pressure of $\sim
10^{-9}$\,mbar in a 3-dimensional magneto-optical trap (3D-MOT). A
moving optical molasses launch accelerates them vertically upwards
to a $\sim 1-$s ballistic trajectory with a temperature of
1.2-2$\,\mu$K. Raman sideband cooling in a co-moving optical
lattice results in $\sim 3\times 10^8$ atoms in the $F=3, m_F=3$
state at a (3D) temperature of 150\,nK
that form a cloud of roughly 3\,mm$^2$ area \cite{Treutlein}. %The
%corresponding source brightness is $2\times 10^22/$m$^5$s$^{-2}$.
A sudden change in the magnetic field followed by a 120-$\mu$s
microwave pulse transfers $\sim 20\%$ of them into the $F=4,
m_F=0$ state. Atoms left over in the $F=3$ state are then cleared
away using a resonant laser pulse. A solenoid generates a small
magnetic bias field to set the quantization axis.

\begin{figure}[t]
\centering {\hspace{-0.5cm}
\epsfig{file=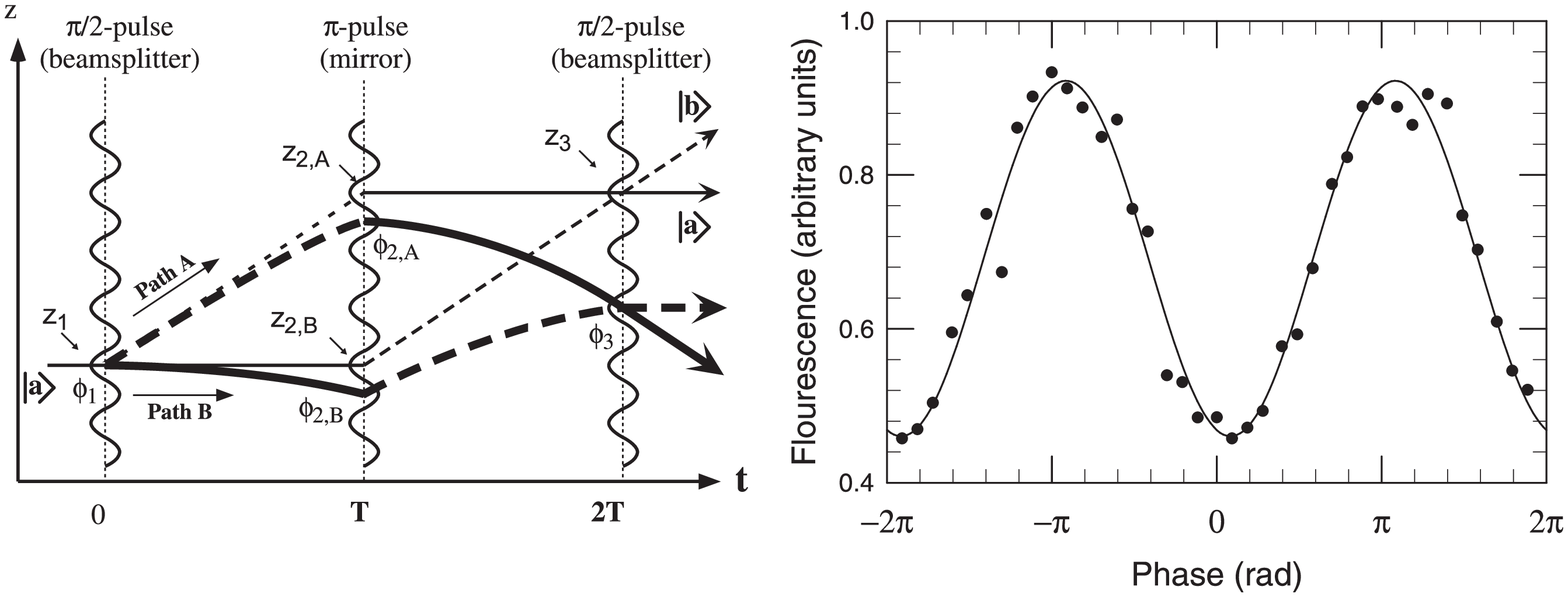,width=0.5\textwidth}}
\caption{Left: Schematic of a Mach Zehnder atom interferometer. In
our experiment, a $\pi$ pulse takes about $100\,\mu$s. Right:
typical fringe obtained in our experiment \label{MZ}}
\end{figure}

Afterwards, a `$\pi/2$' pulse of counterpropagating laser beams,
overlapped with the trajectory of the atoms,  transfers the atoms
into a superposition of the $F=3$ and $F=4$ hyperfine ground
states by a two-photon Raman transition (Fig. \ref{MZ}, left).
These states move vertically relative to each other because of the
momentum of two photons transferred by the laser radiation. After
a time $T\simeq 0.4$\,s, a `$\pi$' pulse interchanges the $F=3$
and 4 states, which afterwards move towards each other. After
another $T$, a final $\pi/2$ pulse recombines the paths to form an
interferometer.

The off-resonant Raman pulses are generated by two extended cavity
diode lasers based on 100-mW laser diodes SDL-5411. The first is
frequency stabilized ('locked') by Doppler-free frequency
modulation (FM) spectroscopy to a Cs vapor cell. It arrives at the
interferometer with a detuning of -1030\,MHz from the $6S_{1/2},
F=3\rightarrow 6P_{3/2}, F'=4$ transition (at 852\,nm wavelength)
in Cs. The second one is phase locked to the first one with a
frequency difference close to the hyperfine splitting of $\simeq
9192$\,MHz, referenced to a LORAN-C frequency standard. 20\,mW of
each laser are transmitted to the experiment via a common
single-mode, polarization maintaining optical fiber. The beams are
then switched and intensity-controlled by an acousto-optical
modulator (Isomet 1205), expanded to about 2.5\,cm, and pass the
vacuum chamber with linear polarization. Retro-reflection on top
of it with two passes through a quarter-wave retardation plate
forms a lin$\perp$lin polarized counterpropagation geometry.

The matter waves in both interferometer paths acquire a phase
difference $\phi$. The contribution of the free evolution, given
by the classical action $S_{Cl}/\hbar$ vanishes for a constant
$g$. However gravity shifts downwards the location at which the
paths interact with the light by $\Delta z= -gt^2/2$, where $t$
denotes time and $z$ the vertical coordinate. This gives rise to a
phase difference (assuming the UFF) \cite{Peters}
\begin{equation}\label{phase}
\phi=k_{\rm eff}gT^2-\phi_L,
\end{equation}
where $\phi_L=\phi_1-2\phi_2+\phi_3$ is given by the phases
$\phi_{1-3}$ of the laser fields at $z=0$. To high accuracy, the
laser radiation can be modelled as a plane wave, which results in
an effective wavevector $k_{\rm eff}=2k=2\omega/c$. For our
experiment, $\phi \simeq 2.3 \times 10^{7}$\,rad. To measure it,
we adjust $\phi_L$ such that $\phi=0$, which corresponds to the
center of the interference pattern. This can be done by using
$\phi_L=rT^2$, i.e. by ramping the difference frequency at a rate
$r$ or a step-wise approximation of such a ramp.

%\subsection{Fountain}

%\subsection{Interferometer}

Both the $F=3$ and 4 interferometer outputs are detected by
flourescence detection with a Hamamatsu R943-02 photomultiplier
tube (PMT). Normalization of the signals takes out atom number
variations.

Fig. \ref{MZ} (right) shows a typical fringe with a pulse
separation time of $T=400\,$ms, taken with 40 launches that take
75\,s total. A sinewave-fit has a phase uncertainty of 0.031\,rad,
and determines $g$ to an uncertainty of $\sim 1.3\times
10^{-9}$\,g. This corresponds to $11\times 10^{-9}\,g/\sqrt{\rm
Hz}$. An improved short-term resolution of $8\times
10^{-9}\,g/\sqrt{\rm Hz}$ can be reached by taking data at the
50\% points of the fringes only. However, as this method is more
sensitive to systematic effects such as drift of the PMT
sensitivity \cite{Peters}, this approach was not followed. Our
resolution is more than 3 times better than the best previously
reported \cite{Peters}. It also surpasses the best classical
absolute gravimeter, the FG-5 falling corner cube gravimeter, by a
factor of about 20.\\

%\begin{figure}
%\centering
%\epsfig{file=histogram97etgTPXO.eps,width=0.5\textwidth}
%\caption{...}
%\end{figure}

%\section{Data analysis}

The notion that gravity might depend on the direction of the
separation $\vec r$ could be described in very simple terms. For
this work, however, we want to use a model that is as general as
possible on the one hand and compatible with accepted principles
that underlie the standard model and gravitational theory on the
other hand. Two such models suggest themselves, Nordtvedt's
anisotropic universe model \cite{Nordtvedt76} and the standard
model extension (SME) \cite{Kostelecky,BaileyKostelecky}. The SME
starts from a Lagrangian formulation of the standard model and
gravity, adding general Lorentz violating terms that can be formed
from the fields and tensors. For the gravitational fields present
on Earth, a post-Newtonian approximation is justified. For two
masses $M$ and $m$, separated by $\vec r$, where $M$ is assumed to
be at rest, the Lagrangian for the gravitational interaction in
the SME is \cite{BaileyKostelecky}
\begin{eqnarray}
\mathcal L=\frac12  m v^2+
G\frac{Mm}{2r}\left(2+3 \bar s^{00}\right. \nonumber \\
\left.+ \bar s^{jk} \hat r^j\hat r^k -3\bar s^{0j} v^j-\bar
s^{0j}\hat r^j v^k \hat r^k\right).
\end{eqnarray}
The indices $j,k$ denote the spatial coordinates, $\vec v$ the
relative velocity, and $\hat r=\vec r/r$. $\bar s^{\mu\nu}=\bar
s^{\nu\mu}$ specifies Lorentz violation in gravity. The two-body
Lagrangian of the anisotropic universe model is similar, but $\bar
s^{00}=0$ and the coefficients of the $v^j$ and the $\hat r^j v^k
\hat r^k$ terms are independent of each other. The equation of
motion (simplified by using $v\ll 1$ and neglecting constant as
well as horizontal accelerations) reads
\begin{equation}
\ddot r^l+\hat r^l\frac{GM}{2r^2}(2+\bar s^{jk}\hat r^j\hat r^k)=0
\end{equation}
where the coefficient of $\hat r^l$ gives the modified
acceleration of free fall.

One outstanding feature of atom interferometry is the relative
simplicity of the underlying theoretical assumptions, which can be
traced to its relying on non-relativistic single-particle effects.
This allows us to analyze the experiment without assuming the LLI
of the Maxwell sector. We therefore determine $k_{\rm eff}$ in Eq.
(\ref{phase}) from the dispersion relation for photons having a
frequency of $\omega_0$ of the SME (neglecting Lorentz-violating
birefringence, which astrophysics experiments bound to $<
10^{-37}$ \cite{KosteleckyMewes2006}, and a constant)
\cite{KosteleckyMewesPRD}
\begin{equation}
k=\omega_0 \left[1-\frac12 (k_F)^{\alpha j \alpha k}\hat k_j\hat
k_k-(k_F)^{\alpha 0 \alpha j} \hat k_j\right],
\end{equation}
where $k_F$ specifies Lorentz violation in the Maxwell sector. As
$\vec k_{\rm eff}=\vec k_1-\vec k_2$, where $\vec k\simeq \vec k_1
\simeq -\vec k_2$, the last term cancels out. In our experiment
the beams are vertical, $\hat k=\vec k/k=\hat r$. Thus, Eq.
(\ref{phase}) reads
\begin{eqnarray}
\phi=2k_0 g_0 \left[1+\frac 12\sigma^{jk}\hat r^j\hat
r^k\right]T^2-\phi_L.
\end{eqnarray}
where $g_0=GM/r^2$ and $k_0=\omega_0/c_0$ \cite{remark}. Thus, the
measured anisotropy is given by $\sigma^{jk}=\bar s^{jk}-
(k_F)^{\alpha j \alpha k}$. Various definitions of coordinates and
fields can still be made, that could be chosen to yield
$(k_F)^{\alpha j \alpha k}=0$.

By coordinate transformations from an inertial sun-centered
celestial equatorial frame (denoted with capital indices $J,K$)
into the laboratory frame on Earth \cite{BaileyKostelecky} we
obtain the time-dependence
\begin{equation}\label{Fourierseries} \frac{\delta
g}{g_0}= \sum_m C_m\cos(\om_m t+\phi_m)+D_m\sin(\om_m t+\phi_m)
\end{equation}
of the $g$-modulations. The coefficients $C_m, D_m$ for the six
frequencies $\om,2\om,\om\pm \Om,2\om\pm \Om$ are functions of the
components of $\sigma^{\mu\nu}$, of Earth's orbital velocity
$v/c\simeq 10^{-4}$, and the frequencies of Earth's orbit
$\Omega=2\pi/(1$\,y) and rotation $\om\simeq 2\pi/(23.93$\,h).

\begin{figure}[t]
\centering \epsfig{file=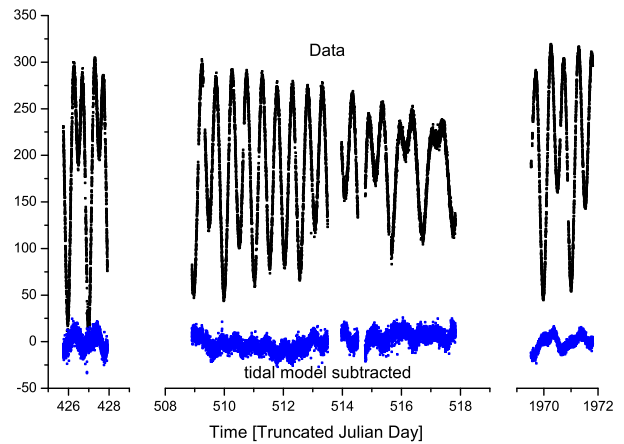, width=0.45\textwidth}
\caption{Data in $10^{-9}g$. Each point represents a 60-s scan of
one fringe (75-s after TJD1900). \label{alldata}}
\end{figure}

For bounding post-Newtonian gravity, we use $\sim$60\,h of data
taken with this setup, as well as a $\sim 60$\,h and a $\sim$10\,d
run reported previously \cite{Peters}, see Fig. \ref{alldata}.
Periodic changes having an amplitude of around
$100\,\mu$gal$\simeq 10^{-7}g$ are due to tides. Subtraction of a
Newtonian model based on the relative positions of the Sun, the
Moon, and the planets \cite{Tamura,Wenzel} yields the graph shown
at the bottom of Fig. \ref{alldata}. More sophisticated tidal
models are available \cite{Bos} that take into account ocean
loading and local effects. However, such models typically rely, in
part, on fitting $g-$observations and are thus not suitable for
our purpose of comparing to a Newtonian model.

The combined data spans about 1500\,d, but fragmented into three
relatively short segments. A Fourier analysis yields the
components given in Tab. \ref{res}. The fragmentation of the data
leads to significant overlap, as given by a covariance matrix
${\rm cov}$. To remove the overlap, we form linear combinations
using ${\rm cov}^{-1}$, see Tab. \ref{res}. The error in this
estimate, given as the geometric sum of the errors entering the
linear combinations, increases in this process.

\begin{table}
\centering \caption{\label{res} Results. Last column gives parts
in $10^{9}$ for $\sigma^{JK}$ and parts in $10^{-5}$ for
$\sigma^{TJ}$.}
\begin{tabular}{cccc}\hline
Comp. & Measured & Disentangled & Result \\ & $10^{-9}$ &
$10^{-9}$ &  \\
\hline
$C_{2\om}$ & 0.342(88) & -0.44(17) & $\sigma^{XX}-\sigma^{YY} = -5.6(2.1)$ \\
$D_{2\om}$ & -0.942(89) & -0.02(19) & $\sigma^{XY} = -0.09(79)$ \\
$C_{\om}$ & 3.668(88) & 3.1(8.8) & $\sigma^{XZ}=-13(37)$ \\
$D_{\om}$ & -0.267(85) & 14.8(9.0) & $\sigma^{YZ} = -61(38)$ \\
$C_{2\om+\Om}$ & -1.378(87) & -1.11(65) & $\sigma^{TY}=172(100)$ \\
$D_{2\om+\Om}$ & -1.051(89) & -1.08(68) & $\sigma^{TX}=-167(104)$ \\
$C_{2\om-\Om}$ & 1.438(89) & -0.30(66) & $\sigma^{TY}=-2.0(4.4)$ \\
$D_{2\om-\Om}$ & 0.536(88) & 0.82(67) & $\sigma^{TX}=5.4(4.5)$ \\
$C_{\om+\Om}$ & 0.647(94) & -12.4(6.2) & $\sigma^{TX}=258(129)$ \\
$D_{\om+\Om}$ &  -2.020(82) & -3.63(6.2) & $\sigma^{TY}-0.21\sigma^{TZ}=76(130)$ \\
$C_{\om-\Om}$ & 1.610(82) & 9.56(6.3) & $\sigma^{TX}=-200(130)$ \\
$D_{\om-\Om}$ & 2.840(92) & 0.11(6.1) & $\sigma^{TZ}+0.21\sigma^{TY}=-00(26)$\\
\hline
\end{tabular}
\end{table}

Comparing the modulations of $g$ given by Eq.
(\ref{Fourierseries}) to the measurement, we obtain the estimates
listed in the fourth column of Tab. \ref{res}. $\sigma^{TZ}$ is
measured as a linear combination with $\sigma^{TY}$, into which we
insert $\sigma^{TY}$ as previously determined. Some components are
multiply determined and could be combined to a weighted average,
but in all cases one limit strongly outweighs the others. Our
final results are (parts in $10^9$)
\begin{eqnarray}
\sigma^{XX}-\sigma^{YY} = -5.6(2.1), \quad \sigma^{XY} =
-0.09(79), \nonumber \\ \sigma^{XZ} =-13(37), \quad \sigma^{YZ} =
-61(38)
\end{eqnarray}
and (parts in $10^5$)
\begin{eqnarray}
\sigma^{TY}=-2.0(4.4), \quad \sigma^{TX}&=&5.4(4.5),\nonumber
\\ \sigma^{TZ}=1.1(26).
\end{eqnarray}

In this letter, we have reported three types of results: First, a
gravimeter based on cold atoms, which uses a pulse separation of
$T=400\,$ms and a bright source of Cs atoms using Raman sideband
cooling in an optical lattice to reach a resolution of
$(8-11)\times 10^{-9}\,g/\sqrt{\rm Hz}$. Second, we analyze the
expected modulation of the local gravitational acceleration
apparent in this experiment as a result of Lorentz violation in
both post-Newtonian gravity and electromagnetism. Third, our test
of the isotropy of post-Newtonian gravity bounds four combinations
of $\sigma^{JK}$ to the $10^{-9}$ level and the three
$\sigma^{0J}$ to the $10^{-5}$ level. Whereas most tests of
gravity are astrophysical in nature \cite{Will71}, ours is a
laboratory experiment, which offers reproducibility and superior
control over relevant parameters.

A previous order-of-magnitude limit $|\bar s^{JK}|\leq 4\times
10^{-9}$ exists, translated \cite{BaileyKostelecky} from the
anisotropic universe bounds due to Nordtvedt \cite{Nordtvedt76}.
No such previous limits on the $\bar s^{TJ}$ are known to us. A
forthcoming publication derives bounds on $\bar s$ from 34 years
of lunar laser ranging (LLR) data that complement our laboratory
bounds \cite{Battat}. We note, however, that ours is the first
experiment where the simultaneous influence of the
non-gravitational and gravitational effects are understood
quantitatively and which accordingly states combined bounds. For
other experiments, these influences are not understood at present.
Moreover, the results differ vastly in the orbit (if one can think
of the atoms' trajectory as an orbit) and quantum-mechanical
nature of the test masses. This is interesting, as quantum gravity
might conceivably involve phenomena that couple to coherent
quantum states but not classical objects.

Future bounds may be found by use of torsion balances, $g$-data
that is routinely taken in geophysical research, or the gravity
probe-B satellite. It is also interesting to study horizontal
interferometer geometries, as they might offer suppression of
tidal influences, which is the main factor limiting our
resolution. In addition, lifting our assumption that UFF is valid,
our data could be analyzed for bounds on $a$- and $c-$ type SME
matter coefficients \cite{KosteleckyTasson}. We remark that
gravity yields space-time varying contributions to $k_F$ related
to Nambu-Goldstone modes \cite{Bluhm}; our analysis uses a flat
space-time picture where those are averaged over
\cite{BaileyKostelecky}. This is likely to yield higher-order
corrections that are currently being investigated
\cite{KosteleckyTasson}. The gravimeter itself is still not
limited by any fundamental limits such as quantum projection
noise. With $\sim 10^8$ atoms per launch, a quantum projection
limited gravimeter could reach the $10^{-12}g$ level per launch
and $10^{-14}g$ per day, if other noise sources (notably phase
noise and vibrations) can be controlled. This promises improved
tests of gravity based on atom interferometry, deepening our
understanding of the fundamental principles of Nature.

We thank A. Peters, Q.G. Bailey, J. Overduin, V.A. Kostelecky, and
A. Mulch for valuable contributions. This material is based upon
work supported by the National Science Foundation under Grant No.
0400866, the Air Force Office of Scientific Research, and the
Multi-University Research Initiative.


\begin{thebibliography}{99}
\bibitem{remark} $\omega_0$ is defined by spectroscopy of thermal atoms in a vapor cell; the stochastic
orientation of their spins should suppress any sensitivity of
$\omega$ to anisotropy.
\end{thebibliography}
\end{document}